\documentstyle[aps,preprint,tighten,floats]{revtex}
\begin{document}
\draft
\title{Gravitational waves from eccentric compact binaries: Reduction
in signal-to-noise ratio due to nonoptimal signal processing}
\author{Karl Martel and Eric Poisson}
\address{Department of Physics, University of Guelph, Guelph,
         Ontario, Canada N1G 2W1}
\maketitle
\begin{abstract}

Inspiraling compact binaries have been identified as one of the most
promising sources of gravitational waves for interferometric
detectors. Most of these binaries are expected to have circularized by
the time their gravitational waves enter the instrument's frequency
band. However, the possibility that some of the binaries might still
possess a significant eccentricity is not excluded. We imagine a
situation in which eccentric signals are received by the detector but 
not explicitly searched for in the data analysis, which uses
exclusively circular waveforms as matched filters. We ascertain the
likelihood that these filters, though not optimal, will nevertheless
be successful at capturing the eccentric signals. We do this by
computing the loss in signal-to-noise ratio incurred when searching
for eccentric signals with those nonoptimal filters. We show that for
a binary system of a given total mass, this loss increases with
increasing eccentricity. We show also that for a given eccentricity,
the loss decreases as the total mass is increased.   
\end{abstract}

\section{Introduction and summary} 

Gravitational waves produced during the last few minutes of inspiral
of compact binary systems --- composed of neutron stars
and/or black holes --- are among the most promising for detection by
kilometer-size interferometric detectors such as LIGO, VIRGO, GEO600,
and TAMA. The ongoing construction of these instruments has
prompted many workers to develop strategies to search for eventual
inspiraling-binary signals in the noisy data 
streams \cite{1,2,3,4,5,6,7,8,9,10,11,12,13,14,15,16,17,18}. The main 
idea is to use the well-known technique of matched filtering \cite{19}
to help find a gravitational-wave signal that may be deeply buried in 
detector noise. This technique consists of cross-correlating the
detector output with a set of model waveforms (often
called templates) which depend on a number of parameters
characterizing the source. These parameters are varied until the
signal-to-noise ratio is maximized, and a signal is concluded to be
present --- within a certain degree of confidence --- if the
signal-to-noise ratio exceeds a certain threshold. Within the class of
linear filtering methods, matched filtering is known to be optimal
\cite{19} when the model waveform which maximizes the signal-to-noise
ratio is an accurate representation of the actual signal. The
development of templates that are sufficiently accurate for the
detection of inspiraling-binary signals is currently a very active
area of research.     

In this continuing effort to generate accurate templates, it is
usually assumed that the binary system moves on a sequence of circular
orbits, with the orbital radius decreasing slowly as a result of
radiation reaction. For most compact binaries, this 
{\it quasi-circular} approximation is well justified, because
radiation reaction decreases the orbital eccentricity as the system
evolves toward smaller orbits \cite{20}. If $a$ and $e$ respectively
denote the binary's semi-major axis and eccentricity, then radiation
reaction changes these quantities according to the approximate
relation $e/e_0 \simeq (a/a_0)^{19/12}$; $a_0$ and $e_0$ are the
initial values \cite{20}. This relation implies that a reduction in
semi-major axis by a factor of 2 comes with a reduction in
eccentricity by a factor of approximately 3. Thus, radiation reaction
quickly circularizes the orbit, and any long-lived compact binary, of
the sort produced when an ordinary (but massive) binary star reaches
the endpoint of stellar evolution, will have become circular by the
time its gravitational waves become measurable. (To be
measurable, the waves must have a frequency that lies within the
detector's frequency band, between 40 Hz and 1000 Hz for the initial
LIGO detector. The dominant contribution to the waves comes at twice
the orbital frequency. Thus, the waves enter the relevant frequency
band when the orbital frequency becomes larger than 20 Hz. For a
binary system consisting of two neutron stars, each of 1.4 solar
masses, this corresponds to an orbital radius of 290 km and an orbital
velocity of $0.12 c$, where $c$ is the speed of light.) 

While probably most binaries will have become circular by the time
their gravitational waves enter the frequency band of interferometric
detectors, the possibility that some binaries might still have a
significant eccentricity is not excluded. The formation scenario
for these eccentric binaries is different from what was considered in
the preceding paragraph. We may imagine that in a densely populated
region of the universe, such as the core of a globular cluster or the
nucleus of a galaxy, a significant number of compact binaries are
produced by two- and three-body processes involving initially isolated
compact objects. Such a scenario may play an important
role in the formation of massive black holes in galactic 
nuclei \cite{21,22}, and it might well provide an interesting number
of gravitational-wave sources. Because some of the resulting binaries
will not have time to eliminate their eccentricities before their
gravitational waves enter the relevant frequency band, it is
conceivable that eccentric compact binaries will make up a sizable
fraction of the total number of gravitational-wave sources.

Admitting this possibility, the issue arises as to how this might
affect the data-analysis strategies. While it is possible to
include the eccentricity in the list of template parameters, doing
so would significantly increase the number of templates required to
search for signals \cite{11,17}. This would come with an increased
computational burden associated with data processing, and a higher 
setting for the threshold value of the signal-to-noise ratio. On the
other hand, using only quasi-circular templates might prevent us from
detecting a potentially interesting number of sources, those
corresponding to eccentric binaries. 

To settle the issue requires information from two fronts. First, we
need a reliable estimate of the number of eccentric binaries that
might be measured by interferometric detectors in the course of a
year. Second, we need to compute the loss of event rate that is
incurred when searching for eccentric signals using nonoptimal,
circular templates. On the basis of this information we could decide 
whether an optimal search for eccentric binaries would be worthwhile,
even at the price of a higher threshold and increased computational
burden, or whether the circular templates would be sufficiently
effective at capturing most of the signals, both circular and
eccentric.  

Our purpose in this paper is to examine the second question, namely,
to estimate the loss of event rate incurred when searching for
eccentric signals using a nonoptimal set of circular templates. Our
main tool in this work is Apostolatos' fitting factor (FF) \cite{8},
which measures the degree of optimality of a given set of
templates. More concretely, the fitting factor is the ratio of the
{\it actual} signal-to-noise ratio, obtained when searching for
eccentric signals using circular templates, to the signal-to-noise
ratio that would be obtained if eccentric templates were used
instead. A fitting factor close to unity indicates that the circular
templates are quite effective at capturing an eccentric signal, and
that little would be gained by using an optimal set of templates. On
the other hand, a fitting factor much smaller than unity indicates
that the circular templates do poorly, and that a set of eccentric
templates would be required for a successful search. The loss in event
rate caused by using nonoptimal templates is given in terms
of the fitting factor by $1 - \mbox{FF}^3$.    

The fitting factor is computed as follows \cite{8}. Let $s(t)$ be the
actual gravitational-wave signal corresponding to an eccentric compact 
binary, and let $h(t;\bbox{\theta})$ denote the circular templates,
with the vector $\bbox{\theta}$ representing the template
parameters. We denote the Fourier transform of these functions by
$\tilde{s}(f)$ and $\tilde{h}(f;\bbox{\theta})$, respectively; 
for any function $a(t)$, $\tilde{a}(f) = \int_{-\infty}^\infty 
a(t) e^{2\pi i f t}\, dt$.  We first define the ambiguity function  
${\cal A}(\bbox{\theta})$ by 
\begin{equation}
{\cal A}(\bbox{\theta}) = \frac{(s|h)}{\sqrt{(s|s)(h|h)}},
\label{1}
\end{equation}
where 
\begin{equation}
(a|b) = 2 \int_0^\infty \frac{ \tilde{a}^*(f) \tilde{b}(f) 
+ \tilde{a}(f) \tilde{b}^*(f) }{S_n(f)}\, df
\label{2}
\end{equation}
is the natural matched-filtering inner product. Here, an asterisk
denotes complex conjugation and $S_n(f)$ is the (one-sided) spectral
density of the detector noise. For the purpose of this work we choose
a noise curve that roughly mimics the expected noise spectrum of the
initial LIGO detector, and set \cite{5}
\begin{equation}
S_n(f) = S_0 \Bigl[ (f_0/f)^4 + 2 + 2(f/f_0)^2 \Bigr]
\label{3}
\end{equation}
for $f > 40\ \mbox{Hz}$, with $f_0 = 200\ \mbox{Hz}$. The value of
$S_0$ is irrelevant for our purposes, and we take $S_n(f)$ to be
infinite below 40 Hz. The fitting factor is the maximum value of the
ambiguity function,
\begin{equation}
\mbox{FF} = \max_{\bbox{\theta}}\, {\cal A}(\bbox{\theta}),
\label{4}
\end{equation}
where the maximization is over all possible choices of source
parameters.  

Expressions for $s(t)$ and $\tilde{h}(f; \bbox{\theta})$ will be
provided in Sec.~II and III, respectively. These expressions are
based upon the rather severe assumptions that the binary's orbital
motion is governed by Newtonian gravity, and the adiabatic evolution
of the orbital elements governed by the Einstein quadrupole
formula. Our waveforms will therefore be left at the ``Newtonian''
level, and we shall ignore all post-Newtonian corrections to the
waveforms. While this level of approximation does not
produce a fully realistic waveform, it should nevertheless be
sufficient for the purposes of this investigation. A more accurate
version of our calculations, based on (say) second post-Newtonian
waveforms \cite{23,24}, would be technically more difficult; for
example, the number of template parameters would have to be increased
from 3 to at least 4. In any event, it is doubtful that any of our
conclusions would be invalidated by a more sophisticated treatment; we
shall come back to this point in Sec.~IV.    

The ambiguity function and fitting factor are computed in Sec.~III.   
For a given binary system, characterized by a choice of individual
masses, the fitting factor is a function of the degree of eccentricity
of the orbital motion. Because the eccentricity evolves as the system
emits gravitational waves, we must find a meaningful way of
parameterizing this. We proceed as follows. We begin the orbital
evolution just before the gravitational waves enter the detector's
frequency band. Because the waves have a frequency component at three
times the orbital frequency (see Sec.~II), this means that we must
begin when $f_{\rm orb} = 13.3\ \mbox{Hz}$, where $f_{\rm orb} = 
(2\pi M)^{-1} (M/a)^{3/2}$ is the orbital frequency. Here, $M$ denotes
the total mass of the binary system and $a$ is the semi-major axis; we
use geometrized units in which $G=c=1$. Having thus specified $a_0$,
the initial value of the first orbital element, we next choose $e_0$,
the initial value of the eccentricity. Thus, $e_0$ is the binary's
eccentricity at the time the gravitational waves enter the
instrument's frequency band. For a given binary system, we 
take FF to be a function of $e_0$. 

\begin{figure}[t]
\special{hscale=60 vscale=60 hoffset=-20.0 voffset=40.0
         angle=-90.0 psfile=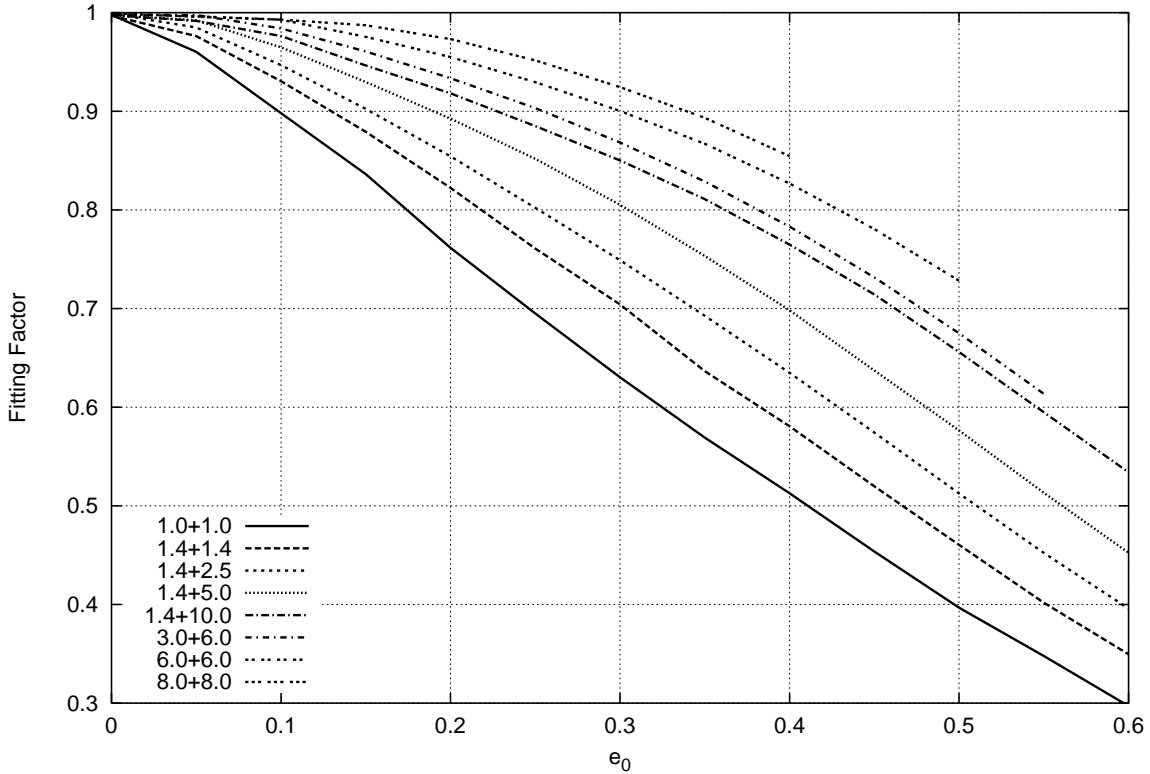}
\vspace*{4.1in}
\caption{The fitting factor as a function of initial eccentricity, 
for selected binary systems. The curves are labelled by the value of
the individual masses, given in units of the solar mass.}
\end{figure}

This function is plotted for selected binaries in Fig.~1. While 
$\mbox{FF} = 1$ when $e_0 = 0$, we see that the fitting factor 
decreases monotonically as $e_0$ increases. This was to be expected: 
As the eccentricity increases, the circular templates become less
effective at capturing the gravitational-wave signal. We also see that
FF decreases faster for small-mass binaries than it does for large-mass
binaries. As we shall see in Sec.~II, this has to do with the fact
that the gravitational-wave signal is shorter for large-mass systems,
so that the circular templates have less of an opportunity to go out
of phase. 

If we select a cutoff $\mbox{FF}|_{\rm min}$ for the fitting factor,
dismissing as inadequate those templates for which $\mbox{FF}
< \mbox{FF}|_{\rm min}$, we see that the range of eccentricities
for which the circular templates {\it are} adequate increases with the
total mass of the binary system. For example, let 
$\mbox{FF}|_{\rm min} = 90\%$, so that the tolerated loss in event
rate is less than 27\%. For a binary system consisting of two 1.4
solar-mass neutron stars, the allowed range of eccentricities is $e_0
\lesssim 0.13$. For a system of two 6.0 solar-mass black holes, the
range is $e_0 \lesssim 0.30$. We conclude that an eccentric binary
system has a better chance of being detected if it is more
massive. This conclusion is reinforced by the fact that massive
binaries also emit stronger waves \cite{16}. These observations
summarize the results of our investigation.             

\section{Gravitational waves from eccentric binaries}

The gravitational waves emitted by an eccentric binary system are
calculated to leading order in a post-Newtonian expansion. In this
approximation, the orbital motion is governed by Newtonian gravity, so
that the (relative) orbit is an ordinary Keplerian ellipse. The waves
are computed on the basis of the Einstein quadrupole formula, and the
adiabatic evolution of the orbital elements is incorporated by taking
into account the loss of orbital energy and angular momentum to the
gravitational waves. The waveforms were first calculated by 
Wahlquist \cite{25}, and our expressions below are identical to his
except for a different notation. The evolution of the orbital elements
was first calculated by Peters \cite{20}. Post-Newtonian corrections
to these results were recently obtained by Gopakumar and 
Iyer \cite{24}; their paper contains the foundations
for an improved version of our work.  

We place the orbit in the $x$-$y$ plane of a Cartesian coordinate
system, and orient the major axis along the $x$ axis. The orbital
radius is expressed as
\begin{equation}
r = \frac{p\, M}{1 + e \cos\Phi},
\label{5}
\end{equation}
where $M = m_1 + m_2$ is the total mass, $p$ a dimensionless
semi-latus rectum, $e$ the eccentricity, and $\Phi$ the orbital
phase, obtained by integrating
\begin{equation}
\frac{d\Phi}{dt} = \frac{ (1 + e \cos\Phi)^2 }{ p^{3/2} M }.
\label{6}
\end{equation}
The orbital period is given by 
\begin{equation}
P = 2\pi M \biggl( \frac{p}{1-e^2} \biggr)^{3/2},
\label{7}
\end{equation}
and the semi-major axis is related to $p$ and $e$ by $a =
p M/(1-e^2)$. 

The orbital elements $p$ and $e$ both decrease as a function of time
by virtue of the fact that the gravitational waves remove energy and 
angular momentum from the system. The relevant expressions 
are \cite{20} 
\begin{equation}
\frac{dp}{dt} = -\frac{64}{5}\, \frac{\mu}{M^2}\, 
\frac{(1-e^2)^{3/2}}{p^3}\, \biggl( 1 + \frac{7}{8}\, e^2 \biggr) 
\label{8}
\end{equation}
and
\begin{equation}
\frac{de}{dt} = -\frac{304}{15}\, \frac{\mu}{M^2}\, 
\frac{(1-e^2)^{3/2}}{p^4}\, e\, 
\biggl( 1 + \frac{121}{304}\, e^2 \biggr),
\label{9}
\end{equation}
where $\mu = m_1 m_2/(m_1 + m_2)$ is the reduced mass. Equations
(\ref{6}), (\ref{8}), and (\ref{9}) determine the orbital motion
completely once initial values are provided for $p$, $e$, and
$\Phi$. We shall return to this point below.  

We place the gravitational-wave detector at a distance $R$ from the
source, in a direction characterized by the polar angles $\iota$ and
$\beta$ relative to the Cartesian frame \cite{1}. If we choose
$\hat{\iota}$ and $\hat{\beta}$ as polarization axes, the two
fundamental polarizations of the gravitational waves are \cite{25} 
\begin{eqnarray}
s_+ &=& -\frac{\mu}{p R} \Biggl\{ \biggl[ 2\cos(2\Phi-2\beta) 
  + \frac{5e}{2}\, \cos(\Phi-2\beta) 
  + \frac{e}{2}\, \cos(3\Phi - 2\beta) 
  + e^2 \cos(2\beta) \biggr] (1+\cos^2\iota) 
\nonumber \\ & & \mbox{}
  + \Bigl[ e\cos(\Phi) + e^2 \Bigr] \sin^2\iota \Biggr\}
\label{10}
\end{eqnarray}
and
\begin{equation}
s_\times = -\frac{\mu}{p R} \Bigl[ 4\sin(2\Phi-2\beta) 
   + 5e\sin(\Phi-2\beta) + e\sin(3\Phi - 2\beta)  
   - 2e^2 \sin(2\beta) \Bigr] \cos\iota.
\label{11}
\end{equation}
Once Eqs.~(\ref{6}), (\ref{8}), and (\ref{9}) have been integrated, the
waveforms can easily be computed. Inspection of 
Eqs.~(\ref{10}) and (\ref{11}) reveals that the gravitational waves
can be decomposed into components that oscillate at once, twice, and
three times the orbital frequency $f_{\rm orb} = 1/P$. (The wave
spectrum is actually more complicated than this because the angular
velocity $d\Phi/dt$ is not uniform. Nevertheless, this
decomposition of the waves into three components is still meaningful
and useful.) The detector responds to the linear combination 
$s(t) = f_+ s_+(t) + f_\times s_\times(t)$, where $f_+$ and $f_\times$
are the detector's beam factors \cite{1}. Our calculations are not
sensitive to the numerical value of the beam factors; we choose $f_+ =
1$ and $f_\times = 0$, so that $s(t) = s_+(t)$. Similarly, our
calculations are not sensitive to the numerical value of $\iota$ and
$\beta$; we choose $\iota = \pi/4$ and $\beta = 0$. 

We assume that the gravitational-wave signal began before the waves
entered the frequency band of our detector. In our computations, it is
sufficient to start the signal immediately before it enters this
band. We shall denote by $f_{\rm min}$ the lower end of the
instrument's frequency band; for the initial version of the LIGO
detector, $f_{\rm min} = 40\ \mbox{Hz}$. The signal component that
first enters the band is the one that oscillates at three times the
orbital frequency. We must therefore impose $3f_{\rm orb} < 
f_{\rm min}$ to ensure that our simulated signal begins sufficiently
early; an {\it actual} signal would of course begin much earlier.  

\begin{figure}[t]
\special{hscale=60 vscale=60 hoffset=-20.0 voffset=40.0
         angle=-90.0 psfile=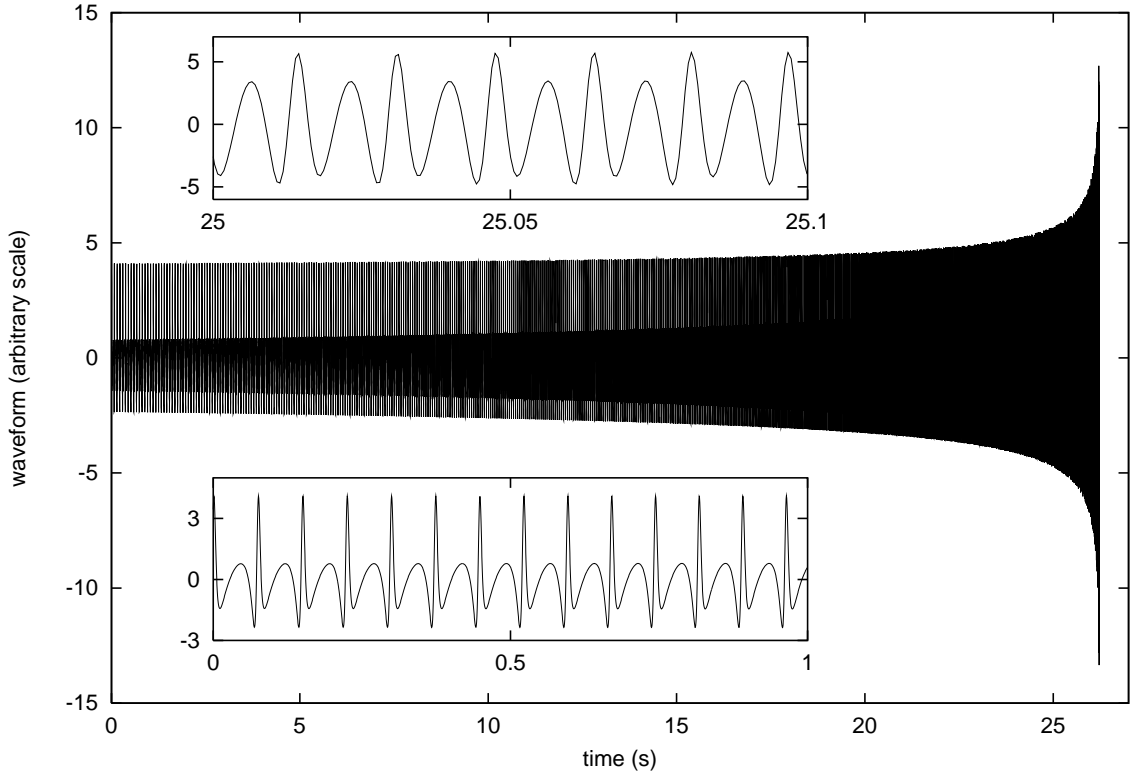}
\vspace*{4.1in}
\caption{Plots of $s(t)$ (up to an overall scaling) for a $1.4 + 1.4$
binary system with initial eccentricity $e_0 = 0.5$. The main figure
shows the waveform for its entire duration. The bottom inset shows the
waveform at early times, when the eccentricity is still large. The top
inset shows the waveform at late times, when the eccentricity is much
reduced. Notice the monotonic increase of both the amplitude and
frequency.}   
\end{figure}

These considerations guide us in the choice of initial values for the
orbital elements. We first select a value for $e_0$, the orbital
eccentricity at the time the gravitational waves enter the detector's
frequency band. We then set $p_0$ equal to 
\begin{equation}
p_0 = \frac{1 - {e_0}^2}{ (2\pi M f_{\rm min} / 3)^{2/3} }
\simeq 180.42 \bigl(1-{e_0}^2 \bigr) 
\biggl( \frac{M_\odot}{M} \biggr)^{2/3},
\label{12}
\end{equation}
so that $f_{\rm orb} = f_{\rm min}/3$ at the initial moment. The
choice of initial value for the orbital phase is inconsequential, and
we simply set $\Phi_0 = 0$. 

In practice, Eqs.~(\ref{6}), (\ref{8}), and (\ref{9}) must be
integrated numerically. We do so with the help of the Bulirsch-Stoer
method ({\it Numerical Recipes} \cite{26}, Sec.~16.4). The functions
$p(t)$, $e(t)$, and $\Phi(t)$ are tabulated (at nonuniform time
intervals selected by the routine's adaptive stepsize controller), 
and cubic spline interpolation ({\it Numerical Recipes}, Sec.~3.3) is
used to evaluate them away from the tabulated points. In this way we
construct the waveform $s(t)$, which is displayed in Fig.~2 for a
specific choice of binary system. 

The Newtonian approximation does not provide a natural cutoff for the
waveforms. Equations (\ref{8}) and (\ref{9}) predict that the orbital
frequency increases all the way to infinity, but this prediction is
unphysical. A relativistic calculation shows instead that the
inspiral proceeds up to a point of instability, from which the two
companions undergo a catastrophic plunge toward each other. The exact
moment of instability is still poorly determined, even in the case of
circular orbits \cite{27,28,29}. As a crude way of cutting off our
waveforms, we simply stop the integration of the orbital equations at
a time $t_{\rm max}$ such that $p \gtrsim 6$. (For a test mass moving
in the gravitational field of a Schwarzschild black hole, $p=6$
designates the innermost stable circular orbit.) Except for the very 
massive binaries, the orbit has essentially become circular by the
time the system reaches the point of instability. 

\begin{table}[t]
\caption{Duration of the gravitational-wave signal (in seconds) as a
function of initial eccentricity $e_0$, for selected binary
systems. The binaries are labelled by the value of their individual 
masses, given in units of the solar mass. The missing entries
correspond to binaries evolving too fast to give a useful
gravitational-wave signal.}
\begin{tabular}{ccccccccc}
$e_0$ & $1.0+1.0$ & $1.4+1.4$ & $1.4+2.5$ & $1.4+5.0$ & $1.4+10$ &
$3.0+6.0$ & $6.0+6.0$ & $8.0+8.0$ \\ 
\hline 
0.00 & 137.97 & 78.75 & 49.24 & 29.04 & 17.60 & 12.65 & 6.96 & 4.31 \\
0.05 & 136.71 & 77.94 & 48.81 & 28.78 & 17.42 & 12.53 & 6.90 & 4.26 \\
0.10 & 133.02 & 75.90 & 47.48 & 28.01 & 16.95 & 12.19 & 6.70 & 4.15 \\
0.15 & 126.99 & 72.49 & 45.32 & 26.73 & 16.19 & 11.64 & 6.41 & 3.96 \\
0.20 & 118.91 & 67.89 & 42.44 & 25.04 & 15.16 & 10.90 & 6.00 & 3.71 \\
0.25 & 110.15 & 62.31 & 38.95 & 22.98 & 13.92 & 10.00 & 5.50 & 3.40 \\
0.30 &  98.03 & 55.95 & 34.99 & 20.63 & 12.50 &  8.99 & 4.94 & 3.06 \\
0.35 &  86.06 & 49.10 & 30.71 & 18.12 & 10.97 &  7.89 & 4.34 & 2.68 \\
0.40 &  73.65 & 42.05 & 26.29 & 15.50 &  9.38 &  6.75 & 3.71 & 2.29 \\
0.45 &  61.28 & 34.97 & 21.87 & 12.89 &  7.81 &  5.62 & 3.08 &  --  \\
0.50 &  49.37 & 28.18 & 17.63 & 10.39 &  6.28 &  4.52 & 2.49 &  --  \\
0.55 &  38.31 & 21.86 & 13.67 &  8.06 &  4.88 &  3.51 &  --  &  --  \\
0.60 &  28.41 & 16.21 & 10.15 &  5.97 &  3.61 &   --  &  --  &  --  \\
0.65 &  19.95 & 11.38 &  7.12 &  --   &  --   &   --  &  --  &  --  \\
\end{tabular}
\end{table}

The values of $t_{\rm max}$ for selected binary systems are listed in
Table I. Since we start the integration at $t = 0$, this corresponds
to the total duration of the signal, from the time it enters the 
instrument's frequency band to the estimated point of instability. The
Table reveals two important trends. First, for a given selection of
masses, the signal duration {\it decreases} with increasing initial
eccentricity. Second, for a given initial eccentricity, the signal
duration {\it decreases} as the mass of the system increases. Both
trends can be understood by combining Eq.~(\ref{8}), evaluated at
$t=0$, with Eq.~(\ref{12}). This gives $|dp/dt|_0 \propto \mu 
{f_{\rm min}}^2 (1-{e_0}^2)^{-3/2} (1+\frac{7}{8}{e_0}^2)$, which
states that increasing either $e_0$ or $\mu$ produces a larger initial
$|dp/dt|$, and therefore a faster evolution.  

\section{Nonoptimal processing of eccentric signal} 

We imagine a detection strategy in which it was decided ahead of time
that the expected signals would come from fully circularized
binaries. The strategy, based on matched filtering, employs a bank of
circular templates $h(t;\bbox{\theta})$. Supposing that some of the
signals come from eccentric binaries, we wish to calculate how much
signal-to-noise ratio is lost when searching for them with nonoptimal
filters. As was explained in Sec.~I, this is given by the fitting
factor, defined in Eq.~(\ref{4}).  

We use the Newtonian approximation to construct the templates. They
can be expressed directly in the frequency domain as \cite{1}
\begin{equation}
\tilde{h}(f;\bbox{\theta}) = A (f/f_0)^{-7/6} e^{i(2\pi f t_c -
\phi_c)}\, e^{i \psi(f)},
\label{13}
\end{equation}
where $A$ is an amplitude parameter, $t_c$ a time-of-arrival
parameter, and $\phi_c$ a phase-at-arrival parameter. Also, 
$f_0$ was defined in Eq.~(\ref{3}), and 
\begin{equation}
\psi(f) = \frac{3}{128}\, (\pi {\cal M} f)^{-5/3},
\label{14}
\end{equation}
where $\cal M$ is the chirp-mass parameter. Because $A$ plays no role
in the calculation of the fitting factor, we set $A=1$ and take
$\bbox{\theta} = (t_c,\phi_c,{\cal M})$. If the binary system were
truly circular, then these parameters would have a clear physical
meaning; for example, the chirp mass would be given in terms of the
individual masses by \cite{1}
\begin{equation}
{\cal M}_{\rm actual} \equiv \frac{(m_1 m_2)^{3/5}}{(m_1+m_2)^{1/5}}.
\label{15}
\end{equation}
Our binaries, however, are not circular, and we shall treat $t_c$,
$\phi_c$, and $\cal M$ as phenomenological parameters with no direct
physical relevance. The issue at stake here is whether the templates
of Eq.~(\ref{13}) are effective at searching signals from eccentric
binaries.  

The Fourier transform of the circular waveforms $h(t,\bbox{\theta})$
can be calculated accurately on the basis of the stationary-phase
approximation \cite{30}; this calculation leads to Eq.~(\ref{13})
above. This approximation, however, is not useful for eccentric
waveforms, because the instantaneous orbital frequency $d\Phi/dt$ is
not a monotonic function of time. We therefore calculate
$\tilde{s}(f)$ by fast Fourier transform 
({\it Numerical Recipes} \cite{26}, Sec.~12.2). Because
$s(t)$ contains three frequency components (at once, twice, and three
times the orbital frequency), the Fourier transform of an eccentric
signal displays more structure than that of a circular signal. This is
illustrated in Fig.~3, which shows the emergence of the three
frequency components in $|\tilde{s}(f)|$, as well as the interference
between them.  

\begin{figure}[t]
\special{hscale=60 vscale=60 hoffset=-20.0 voffset=40.0
         angle=-90.0 psfile=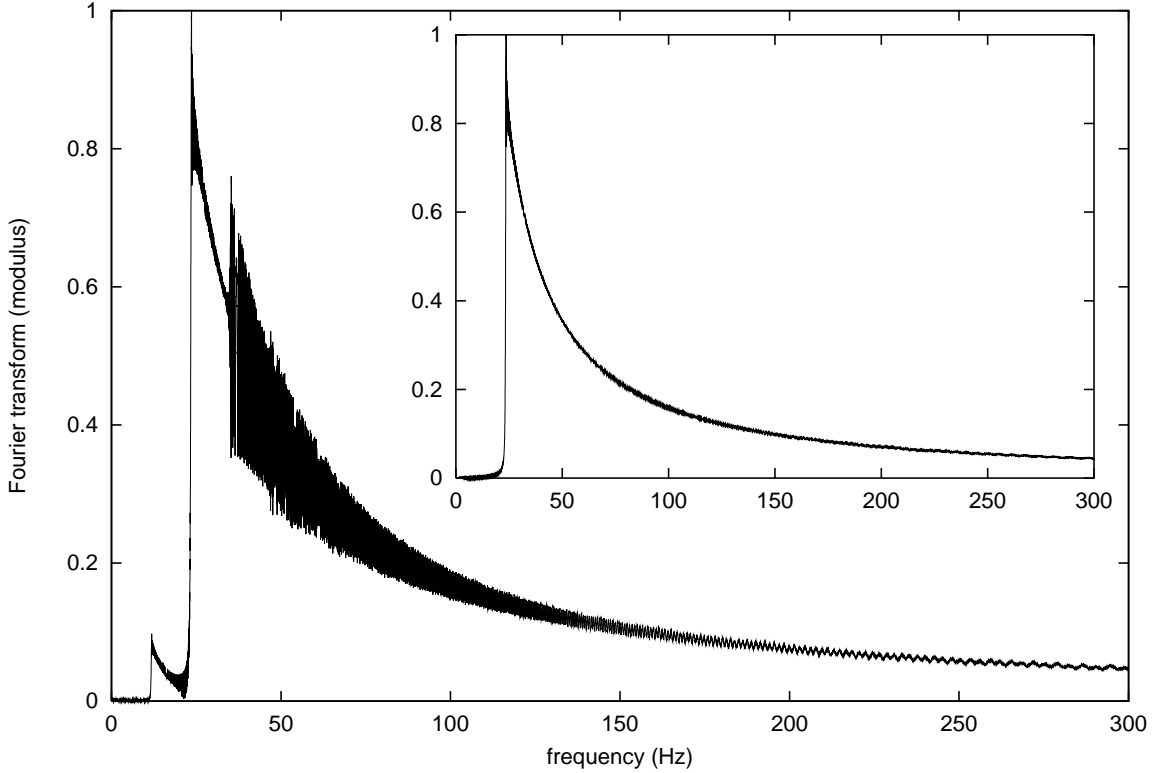}
\vspace*{4.1in}
\caption{Plots of $|\tilde{s}(f)|$ for a $1.4+1.4$ binary system. The
main figure shows the Fourier transform of a signal with initial
eccentricity $e_0 = 0.1$. As explained in the text, the signal is
taken to start abruptly when $f_{\rm orb} = 13.3\ \mbox{Hz}$, which
explains the (artificial) cutoffs at low frequency. The three
components of the gravitational-wave signal, and the interference
between them, are clearly apparent. For comparison, the inset shows
the Fourier transform of a circular signal, which also starts abruptly
when $f_{\rm orb} = 13.3\ \mbox{Hz}$.}  
\end{figure}

We now describe the calculation of the ambiguity function and its
maximization over the template parameters. The overlap integral
between signal and templates can be written as
\begin{equation}
(s|h) = 2 \Bigl[ e^{i\phi_c} {\cal B}(t_c, {\cal M}) +     
        e^{-i\phi_c} {\cal B}^*(t_c, {\cal M}) \Bigr],
\label{16}
\end{equation}
where
\begin{equation}
{\cal B}(t_c, {\cal M}) = \int_0^\infty 
\frac{ (f/f_0)^{-7/6} e^{i\psi} \tilde{s}(f) }{S_n(f)}\, 
e^{-2\pi i f t_c}\, df.
\label{17}
\end{equation}
We note that $(h|h)$ is independent of $\phi_c$ and $t_c$, so that 
maximization of the ambiguity function over these parameters is
achieved by maximizing $(s|h)$ alone. Maximization over $\phi_c$ is 
straightforward, and we obtain
\begin{equation}
\max_{\phi_c}\, (s|h) = 4 |{\cal B}(t_c, {\cal M})|.
\label{18}
\end{equation}
The integral of Eq.~(\ref{17}) is naturally evaluated by fast Fourier
transform. For a given value of $\cal M$, this returns a discretized
version of the function ${\cal B}(t_c)$. Maximization over $t_c$ is
achieved simply by selecting the largest value of $|{\cal B}(t_c)|$,
which we denote $|{\cal B}|_{\rm max}$. These manipulations give us a
reduced ambiguity function,
\begin{equation}
{\cal A}({\cal M}) \equiv 
\max_{t_c,\phi_c}\, {\cal A}(t_c,\phi_c,{\cal M}) =
\frac{ 4 |{\cal B}|_{\rm max} }{ \sqrt{(s|s)(h|h)} },
\label{19}
\end{equation}
in which $(s|s)$ is evaluated as a discrete sum using the fast Fourier
transform of $s(t)$, and $(h|h)$ is evaluated by Romberg integration
({\it Numerical Recipes} \cite{26}, Sec.~4.3).  

The fitting factor is obtained by finding the maximum of 
${\cal A}({\cal M})$. We first sketch this function by evaluating 
the right-hand side of Eq.~(\ref{19}) for several values of 
${\cal M}$. We observe that the maximum always occurs in the interval  
$1.0 < {\cal M}/{\cal M}_{\rm actual} < 1.2$. Having thus bracketed 
the maximum, we find its position by golden-section search 
({\it Numerical Recipes} \cite{26}, Sec.~10.1). 

Table II contains our main results: It lists, for selected binary
systems, the fitting factor as a function of $e_0$, the initial
orbital eccentricity. We estimate the accuracy of our results to be
within 1\%. Plots of the fitting factor were presented in Fig.~1, and
a discussion of our results can be found at the end of Sec.~I. In
Table III we list the values of the ratio 
${\cal M}/{\cal M}_{\rm actual}$ that maximizes 
${\cal A}({\cal M})$. We see that the circular templates attempt to 
compensate for orbital eccentricity by overestimating the chirp-mass
parameter. This was to be expected, because increasing the chirp mass 
produces a faster orbital evolution, which is also an effect of
increasing the eccentricity. 

\begin{table}[t]
\caption{Fitting factor as a function of initial eccentricity $e_0$,
for selected binary systems. Two independent numerical codes were
written to calculate these numbers. The results differ by less than
1\%, and we take this to be an estimate of our numerical accuracy.} 
\begin{tabular}{ccccccccc}
$e_0$ & $1.0+1.0$ & $1.4+1.4$ & $1.4+2.5$ & $1.4+5.0$ & $1.4+10$ &
$3.0+6.0$ & $6.0+6.0$ & $8.0+8.0$ \\ 
\hline 
0.00 & 0.998 & 0.997 & 0.999 & 0.998 & 0.998 & 0.998 & 0.998 & 0.999 \\
0.05 & 0.960 & 0.976 & 0.985 & 0.992 & 0.992 & 0.998 & 0.996 & 0.996 \\
0.10 & 0.898 & 0.931 & 0.947 & 0.965 & 0.976 & 0.984 & 0.993 & 0.993 \\
0.15 & 0.836 & 0.879 & 0.902 & 0.930 & 0.946 & 0.961 & 0.975 & 0.987 \\
0.20 & 0.762 & 0.822 & 0.854 & 0.893 & 0.913 & 0.934 & 0.955 & 0.973 \\
0.25 & 0.695 & 0.761 & 0.802 & 0.852 & 0.885 & 0.903 & 0.930 & 0.952 \\
0.30 & 0.630 & 0.637 & 0.749 & 0.805 & 0.850 & 0.868 & 0.900 & 0.925 \\
0.35 & 0.569 & 0.581 & 0.693 & 0.753 & 0.811 & 0.829 & 0.867 & 0.893 \\
0.40 & 0.513 & 0.520 & 0.635 & 0.698 & 0.765 & 0.783 & 0.827 & 0.854 \\
0.45 & 0.454 & 0.460 & 0.574 & 0.637 & 0.714 & 0.732 & 0.781 &  --   \\
0.50 & 0.397 & 0.402 & 0.513 & 0.576 & 0.656 & 0.675 & 0.728 &  --   \\
0.55 & 0.348 & 0.350 & 0.452 & 0.513 & 0.595 & 0.614 &  --   &  --   \\
0.60 & 0.297 & 0.303 & 0.396 & 0.452 & 0.534 &  --   &  --   &  --   \\
0.65 & 0.257 & 0.231 & 0.344 &   --  &   --  &  --   &  --   &  --    \\
\end{tabular}
\end{table}

\begin{table}[t]
\caption{Value of ${\cal M}/{\cal M}_{\rm actual}$ that maximizes
the reduced ambiguity function.}
\begin{tabular}{ccccccccc}
$e_0$ & $1.0+1.0$ & $1.4+1.4$ & $1.4+2.5$ & $1.4+5.0$ & $1.4+10$ &
$3.0+6.0$ & $6.0+6.0$ & $8.0+8.0$ \\ 
\hline 
0.00 & 1.0000 & 0.9999 & 0.9999 & 0.9999 & 0.9997 & 0.9997 & 0.9994 & 0.9992 \\
0.05 & 1.0007 & 1.0006 & 1.0007 & 1.0007 & 1.0007 & 1.0006 & 1.0004 & 1.0005 \\
0.10 & 1.0012 & 1.0016 & 1.0017 & 1.0022 & 1.0030 & 1.0031 & 1.0033 & 1.0036 \\
0.15 & 1.0024 & 1.0027 & 1.0031 & 1.0039 & 1.0053 & 1.0056 & 1.0076 & 1.0094 \\
0.20 & 1.0037 & 1.0042 & 1.0048 & 1.0060 & 1.0083 & 1.0083 & 1.0113 & 1.0160 \\
0.25 & 1.0059 & 1.0059 & 1.0071 & 1.0087 & 1.0122 & 1.0122 & 1.0165 & 1.0222 \\
0.30 & 1.0088 & 1.0095 & 1.0106 & 1.0121 & 1.0172 & 1.0170 & 1.0228 & 1.0314 \\
0.35 & 1.0125 & 1.0134 & 1.0149 & 1.0167 & 1.0234 & 1.0232 & 1.0312 & 1.0429 \\
0.40 & 1.0182 & 1.0194 & 1.0210 & 1.0223 & 1.0319 & 1.0314 & 1.0418 & 1.0563 \\
0.45 & 1.0260 & 1.0288 & 1.0302 & 1.0307 & 1.0430 & 1.0416 & 1.0562 &  --    \\
0.50 & 1.0404 & 1.0412 & 1.0447 & 1.0466 & 1.0586 & 1.0574 & 1.0774 &  --    \\
0.55 & 1.0598 & 1.0654 & 1.0640 & 1.0680 & 1.0842 & 1.0749 &   --   &  --    \\
0.60 & 1.0986 & 1.0946 & 1.0986 & 1.0976 & 1.1138 &   --   &   --   &  --    \\
0.65 & 1.1508 & 1.1542 & 1.1484 &   --   &   --   &   --   &   --   &  --    \\
\end{tabular}
\end{table}

\section{Conclusion}

\begin{figure}[t]
\special{hscale=60 vscale=60 hoffset=-20.0 voffset=40.0
         angle=-90.0 psfile=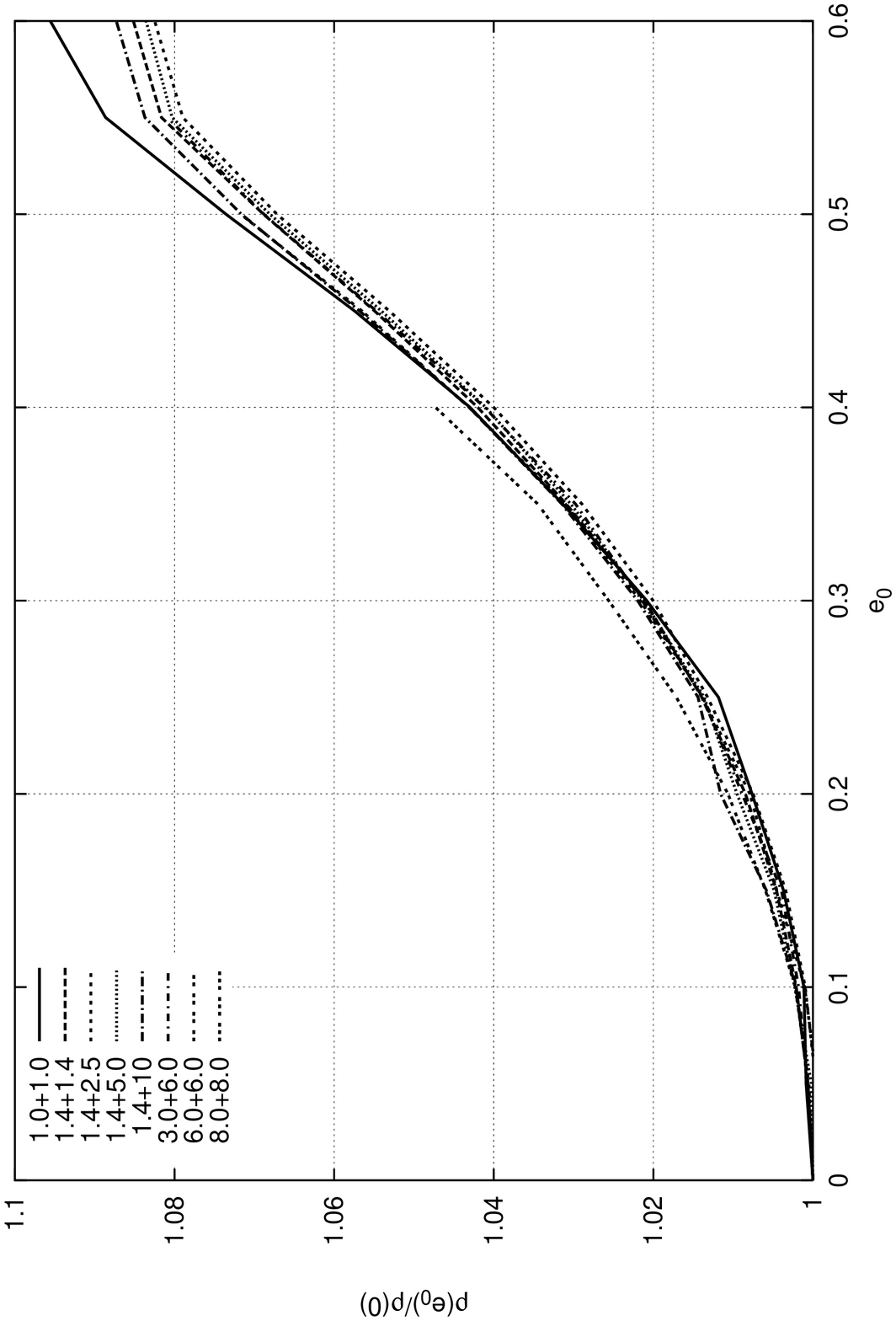}
\vspace*{4.1in}
\caption{The ratio $\rho(e_0)/\rho(0)$ as a function of initial
eccentricity, for selected binary systems. The scatter in the curves
is mostly an artifact of numerical error.} 
\end{figure}

The message of Table II and Fig.~1 is clear: Searching for eccentric
signals with circular templates produces a loss of signal-to-noise
ratio that increases with increasing eccentricity. We recall that the
fitting factor is a measure of this loss of signal-to-noise ratio:
$\mbox{FF} = \mbox{SNR}|_{\rm actual}/\mbox{SNR}|_{\rm optimal}$,
where $\mbox{SNR}|_{\rm actual}$ is the signal-to-noise ratio obtained
with nonoptimal (circular) templates, while 
$\mbox{SNR}|_{\rm optimal}$ is the signal-to-noise ratio that would be
obtained if optimal (eccentric) templates were used instead. This is
given by 
\begin{equation}
\mbox{SNR}|_{\rm optimal} = \sqrt{ (s|s) } \equiv \rho.
\label{20}
\end{equation}
Another important piece of information is how $\rho$ varies with
$e_0$. Because an eccentric signal is essentially a superposition of
three frequency components, while a circular signal contains a single
component at twice the orbital frequency, we expect that $\rho(e_0)$
should increase with increasing eccentricity. Figure 4 confirms this
expectation by showing that the relative increase in
signal-to-noise ratio, measured by $\rho(e_0)/\rho(0)$, is indeed an
increasing function of $e_0$. (This means that an eccentric binary
system located at a distance $R$ from the detector produces
a stronger signal than a circular system at the same distance.) The
increase, however, is modest with respect to the decrease incurred by
nonoptimal filtering. The net effect is still a decrease in
signal-to-noise ratio. 

It would be highly desirable to extend our results to other
gravitational-wave detectors --- such as the advanced version of
LIGO --- and to post-Newtonian waveforms. The advanced LIGO
detector is characterized by a wider frequency band starting at $f =
10\ \mbox{Hz}$; because the circular templates must then stay in phase
with the eccentric signal over a longer period, we would anticipate
lower values for the fitting factor. However, the dependence of the
fitting factor on initial eccentricity and binary mass should be
qualitatively the same. On the other hand, we would anticipate that
post-Newtonian waveforms will produce higher values for the fitting
factor. The reason is that the waveforms now come with additional
parameters, and the added flexibility in the maximization procedure
will likely improve the overlap between signal and templates. An
interesting question is whether the eccentric signals force the
template parameters outside of their natural range. For example, the
dimensionless reduced-mass parameter $\eta = m_1 m_2/(m_1+m_2)^2$ is
physically restricted to the interval $\eta \leq \frac{1}{4}$, but the
value returned by the maximization procedure might exceed this
limit. If this were the case, then a larger number of templates would
be required in a search for eccentric binaries. We hope to return to
these issues in a future publication. 

\section*{Acknowledgments}

This work was supported by the Natural Sciences and Engineering
Research Council of Canada. We thank Scott Hughes and an anonymous
referee for their comments on a previous version of this manuscript.

\end{document}